\documentclass[%
preprint,
 amsmath,amssymb,
 aps,
 prl,
]{revtex4-1}

\usepackage{setspace}
\usepackage{algcompatible}
\usepackage[usenames,dvipsnames]{color}
\usepackage{amsthm}
\newtheorem{definition}{Definition}
\usepackage{graphicx}
\usepackage{dcolumn}
\usepackage{bm}
\usepackage{hyperref}
\usepackage[mathlines]{lineno}

\begin{document}


\title{Graph Compartmentalization}

\author{Matthew J. Denny}
\email{mdenny@polsci.umass.edu}
 \affiliation{422 Thompson Hall, University of Massachusetts Amherst\\ Amherst, MA 01003}

\date{\today}

\begin{abstract}
\singlespacing
This article introduces a concept and measure of graph compartmentalization. This new measure allows for principled comparison between graphs of arbitrary structure, unlike existing measures such as graph modularity. The proposed measure is invariant to graph size and number of groups and can be calculated analytically, facilitating measurement on very large graphs. I also introduce a block model generative process for compartmentalized graphs as a benchmark on which to validate the proposed measure. Simulation results demonstrate improved performance of the new measure over modularity in recovering the degree of compartmentalization of graphs simulated from the generative model. I also explore an application to the measurement of political polarization.
\end{abstract}

\pacs{05}
\maketitle


\section{Introduction}
\label{sec: introduction}
 A number of studies have sought to identify distinct communities in graphs -- using spectral bisection methods \citep{Fiedler1973,Pothen1990}, betweenness centrality \citep{Girvan2002} and modularity \citep{Newman2004, Danon2005, Newman2006, Porter2009a, Mucha2010} among other techniques. The degree to which graphs exhibit separated communities has been found to effect resiliency to congestion and node failure \citep{Callaway2000, Dodds2003}, signal political polarization \citep{Lazer2009}, and characterize sensitive information networks \cite{Coffman2004} and criminal activity \cite{Baker1993}. In particular, studies of community structure have traditionally asked the question: ``for given a graph, what vertex partition contains the most within-group ties"? 
 
This study flips the motivating question on its head, asking instead: ``given a set of community memberships, to what degree is the observed graph characterized by strongly disconnected communities"? Furthermore, this begs a question about the generative process for ties in such a graph: ``did the observed disconnection between communities arise by chance or through a process of preferential edge formation within communities"? 

\begin{definition}
\singlespacing
Let the degree to which a graph is characterized by separation on community membership as a result of a preference for within community edge formation be the \textbf{compartmentalization} of that graph. 
\end{definition}
 
\noindent This study introduces a  measure of graph compartmentalization and a simple generative block model for compartmentalized graphs on which to test this measure. The proposed measure can be seen as a reformulation of modularity that is grounded in expectations about the graph generative process as opposed to the empirical likelihood of edges \citep{Newman2004}. This measure of graph  compartmentalization is then compared against modularity on real and simulated graphs and demonstrates improved performance as a metric of comparison. 

\vspace*{-.2in}
\section{\label{sec:measure}A Measure of Graph Compartmentalization}

A graph that has a compartmentalized structure is one where, given a community membership for each node, a large proportion of edges are sent within communities relative to between communities. While having a high proportion of edges within communities is a necessary identifying feature of a highly compartmentalized graph, it is not sufficient. Consider the case of a graph $(N = 100)$ where only one edge exists, and that edge is between two nodes in the same community. We could take this as evidence that the graph has a highly  compartmentalized structure, but it could also arise with high probability from a generative process without any preference for in-community edge formation.  Now imagine a graph where all edges occur within communities and the graph density $D$ is equal to the maximum possible density that could be attained for that graph with only within-group edges. This constitutes the strongest evidence we can get (without actually knowing the generative process) that the graph arose from a generative process with a perfect preference for within-community edge formation -- a highly compartmentalized graph.


Some studies have sought to compare the compartmentalization of graphs using their modularity as a measure. The modularity of a graph measures the degree to which which edges are concentrated between nodes partitioned into separated groups relative to a random assignment of ties. Following Newman \citep{Newman2004}, for a division of the graph into $L$ distinct communities, define an $L\times L$ matrix $e$ whose $e_{ij}$ component is the proportion of edges in the original graph that connect nodes in group $i$ to those in group $j$. The modularity of the graph is then defined to be:
\begin{equation}
Q = \sum_i e_{ii} - \sum_{ijk} e_{ij}e_{ki} = \text{Tr } \text{\textbf{e}}  - ||\bf{ e^2 }||
\end{equation}
This measure can be maximized to discover communities in an observed graph, but as Newman \citep{Newman2006} notes, it is not intended to qualify graph structure when community membership is known and fixed. Furthermore, $Q$ is not invariant in the number of  or relative size of groups \citep{Danon2005}, and by extension for a fixed number of groups, to graph size. This makes modularity an inappropriate measure for comparison across graphs of arbitrary structure.

I propose a new measure of graph compartmentalization that is related to modularity, but allows for comparison between graphs of arbitrary structure. We begin with a graph $G$ comprised of a set of $N$ nodes with a given vector of group memberships $m = \{m_1 \ldots m_l\}$ (where the value of each node's community membership is $ l \in L$ distinct community  assignments). Let $M$ be a matrix such that $M_{ij} = 1$ if $m_i = m_j$ and zero otherwise. Then we can define $D_M$ as the maximum density the graph could attain with only in-community edges. A set of criteria that a valid measure of graph compartmentalization, $\Upsilon$ must satisfy are listed below. If the proposed measure can be shown to be consistent with these criteria then it will provide a graph size and  community-membership-structure invariant measure of compartmentalization that can facilitate comparison between graphs.
\begin{enumerate}
\singlespacing
\item $\Upsilon$ must be invariant in $N$  and the number and relative size of communities for a constant $D_M$. 
\item $\Upsilon$ must be bounded above and below to give a absolute, comparable measure of compartmentalization across multiple graphs.
\item $\Upsilon$ must only attain its global maximum (minimum) value when $D = D_M$ ($D =1- D_M$) and ties are only present within (between) community. 
\end{enumerate}
Let $A$ be the graph adjacency matrix (with $||A||$ the sum over the adjacency matrix). Then we can define $F$, the fraction of observed edges that occur within-groups as follows:
\begin{equation}
F = \frac{\sum_i \sum_j M_{i,j}A_{i,j}}{||A||}  
\end{equation}
For a given $F$ and $D_M$, we can then define a measure of the compartmentalization of a graph $\Upsilon$ as:
\begin{equation}
\Upsilon = \left[F - D_M\right] \times \begin{cases}\frac{\left[1 - \left( D -D_M \right)^2 \right]}{1 - D_M} \hspace*{.425in} \text{ if } F \geq D_M \\
\frac{\left[1 - \left( D -(1 -D_M) \right)^2\right]}{D_M} \hspace*{.2in}\text{ if } F < D_M
\end{cases}
\end{equation}
The first term, $\left[F - D_M\right]$ bears a strong analogy to the measure of modularity $Q$, as it is the proportion of in-community edges minus the expected proportion of in-community edges if $G$ were generated from the block model described above with $\rho = 0.5$, indicating no preference for within community edge formation (see the middle level plot in Figure \ref{density_proportion_tradeoff 0 100}). The second set of terms function as a relative density correction for this measure so that it is maximized (minimized) when the evidence for compartmentalization (anti-compartmentalization) is maximized. $\Upsilon$ is increasing in $F$ and decreasing in $D$ :

\begin{equation}
\frac{\partial \Upsilon}{\partial F} =  \begin{cases} \frac{\left[1 - \left( D -D_M \right)^2 \right]}{1 - D_M} \hspace*{.425in} \text{ if } F \geq D_M \\
\frac{\left[1 - \left( D -(1 -D_M) \right)^2\right]}{D_M} \hspace*{.2in}\text{ if } F < D_M
\end{cases} \geq 0
\end{equation}
\begin{equation}
\frac{\partial \Upsilon}{\partial D} =  \begin{cases}   \frac{-2 \left[(F +D_M)(D + D_M) \right]}{1 - D_M} \hspace*{.575in} \text{ if } F \geq D_M \\
\frac{ -2 \left[F(1 + D - D_M) + D_M(1 + D)\right]}{D_M} \hspace*{.2in}\text{ if } F < D_M
\end{cases} \leq 0
\end{equation}
\noindent This is consistent with the intuition that more compartmentalized graphs have a higher portion of within-group edges and that more dense graphs are generally less partitioned. This measure also qualifies the strength of our evidence about the relative compartmentalization of the graph. When $F \geq D_M$ we down-weight our evidence $\left[F - D_M\right]$ by its distance from $D = D_M$ and when $F < D_M$ we down-weight by the distance from $D = 1 - D_M$.

Because $\Upsilon$ is normalized by the difference between $D$ and $D_M$ (or 1 - $D_M$) , this measure implicitly assumes that nodes may form atleast as many edges as there are members of their group (out group). This assumption reasonably holds for most commonly studied social networks with groups of less than $\sim 100$ nodes and is necessary to preserve the desired properties of $\Upsilon$ discussed above. However, if it is unreasonable to assume that a node could form edges to all members of its group, a more appropriate normalization would involve dividing the average degree of $G$ by the maximum observed degree. In this formulation we can define $\widetilde{\Upsilon}$ as:
\begin{equation}
\widetilde{\Upsilon} = \left[F - D_M\right] \times \left\{ \frac{\sum_i A_{i,j}}{N\left(\max \sum_i A_{i,j}\right)} \hspace*{.175in} \text{ if } D > 0 ,\text{ } 0 \text{ when } D = 0 \right\}
\end{equation}

\noindent We can see that $\widetilde{\Upsilon}$ = $\Upsilon$ when $D$ = 0 and when $D = D_M$  (or 1 - $D_M$) but the value of $\widetilde{\Upsilon}$ will diverge from $\Upsilon$ especially at high values of $D$. $\widetilde{\Upsilon}$ is also not invariant to the degree distribution of $G$ (which may be theoretically relevant in some applications). As $\Upsilon$ satisfies all of the criteria for a valid measure of compartmentalization laid out above, it is the primary focus of the rest of this paper and investigation of the alternate formulation $\widetilde{\Upsilon}$ is left to future work.

\section{A Generative Block Model for Compartmentalized Graphs}
\label{sec:gen model}

A natural example of a compartmentalized graph is the set of friendship relations between employees in a large company. Employees who work in the same department will have much more interaction with eachother than employees in different departments and therefore be more likely to form friendships. The degree of compartmentalization in friendships in a company is likely to vary with the physical distance between offices of employees in different departments, representing a continuum between low  compartmentalization when all offices in a company share the same space, to very high compartmentalization when different departments are located in different buildings or even different states. An edge formation process consistent with the intuition laid out above can be represented by a block model where whether or not an edge is formed within community is sampled first using a method similar to urn randomization \citep{Wei1988,Schulz2002}, and then the nodes connected by that edge are sampled conditional on whether the edge connects members of the same community.

Let $ \rho \in [0,1]$ be the degree of node preference for edge formation within community such that $\rho = 0$ implies that as long as edges can possibly be formed outside of their group, nodes will choose to do so with probability 1 and $\rho =1$ implies that actors have a perfect preference for within-community edge formation if possible.  Let $D_{in}$ be the density contribution of in-community edges and $D_{out}$ be the density contribution of between-community edges such that $D_{in} + D_{out} = D$, the total density of the graph. Furthermore, let $T = \{t_1 \ldots t_k \}$ be the set of $k$ already existing edges in the graph. Then we can define the probability of an edge forming within-group $\gamma$ as:
\begin{equation}
\gamma = \frac{\left(D_M - D_{in}\right) \rho}{\left(D_M- D_{in}\right) \rho + \left((1- D_M) - D_{out}\right) (1- \rho)}
\end{equation}

\begin{figure}
\caption{\label{fig:generative process graphical model} {\scshape Generative Process}}
\singlespacing
\begin{algorithmic}
\FOR{$k \in K$}
        \STATE{Sample Whether Edge in Community $\sim$ $\gamma(T,\rho, M)$}
        \IF{Edge Within Community}
				\STATE{Sample $S$, $R$ from Shared Community}
				\ELSE
				\STATE{ Sample $S$, $R$ from Different Community}
				\ENDIF
\ENDFOR
\end{algorithmic}
\end{figure}

\begin{figure*}
\singlespacing
\centering
\caption{\label{fig:Simulated_Networks}
Graphs simulated from the generative process with $N = 20$, $T =50$, \textbf{ (a)} : $\rho = 1$,\textbf{ (b)} :  $\rho = 0.85$, \textbf{ (c)} : $\rho = 0$. \bigskip}
\begin{tabular}{ccc}
\textbf{\large(a)} & \textbf{\large(b)} & \textbf{\large(c)}\\
\includegraphics[scale=.4]{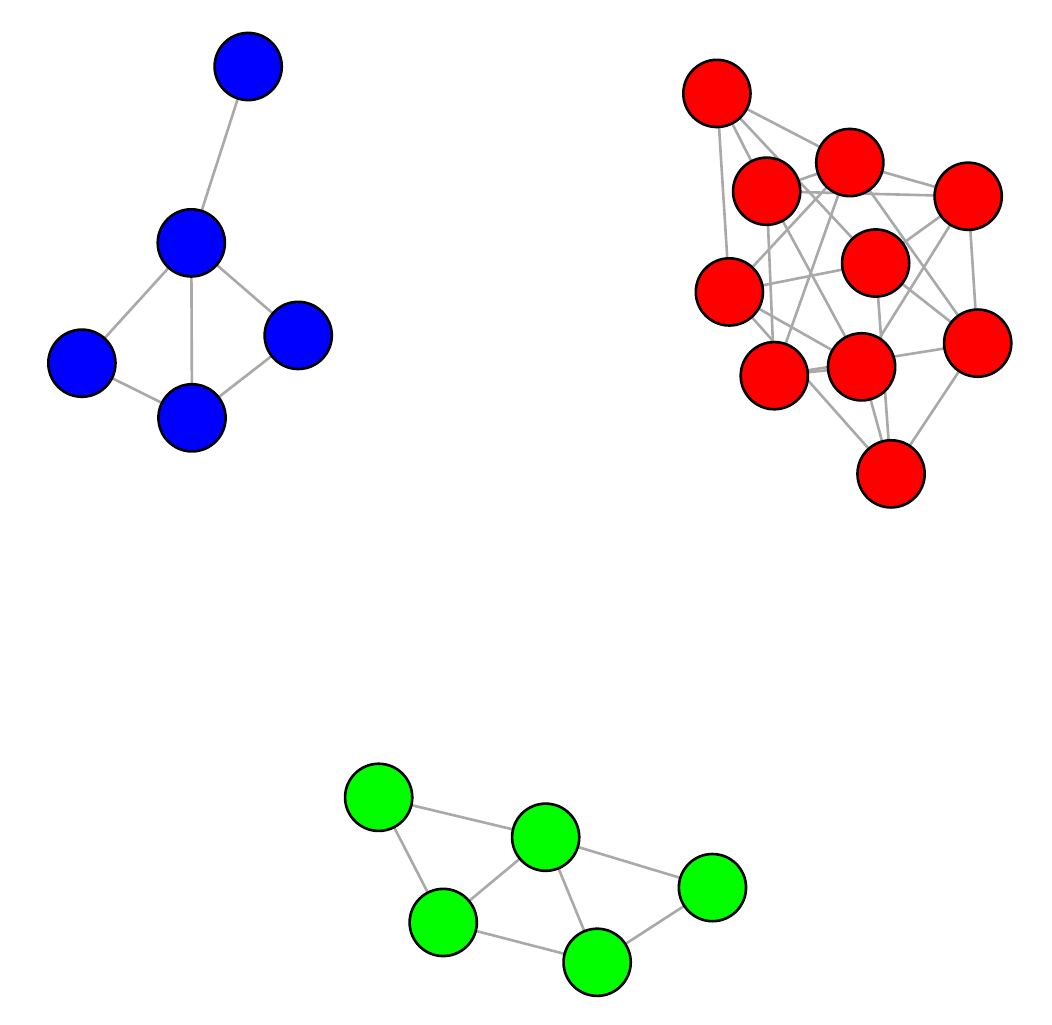} \hspace*{.4in}&
\includegraphics[scale=.4]{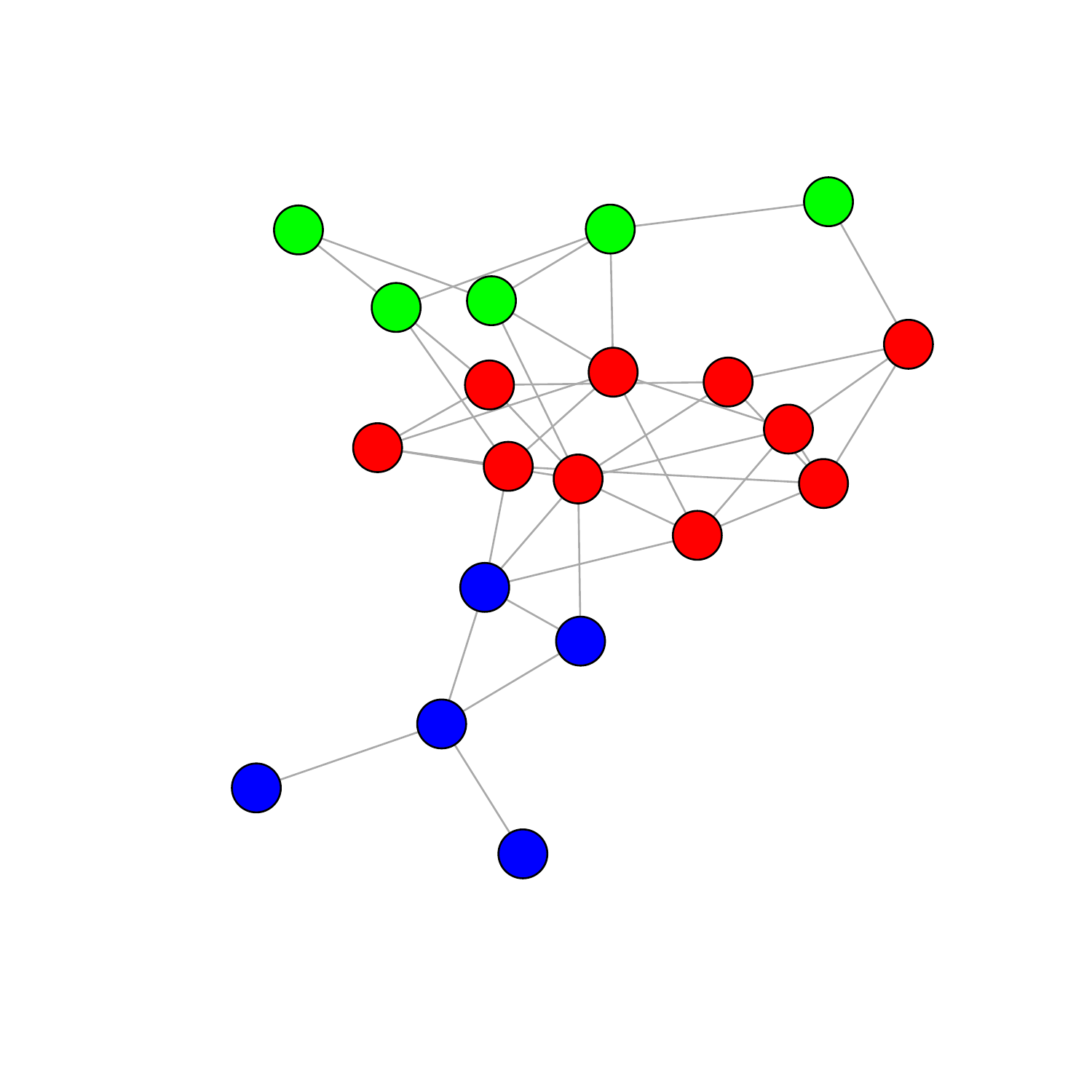} \hspace*{.4in}&
\includegraphics[scale=.4]{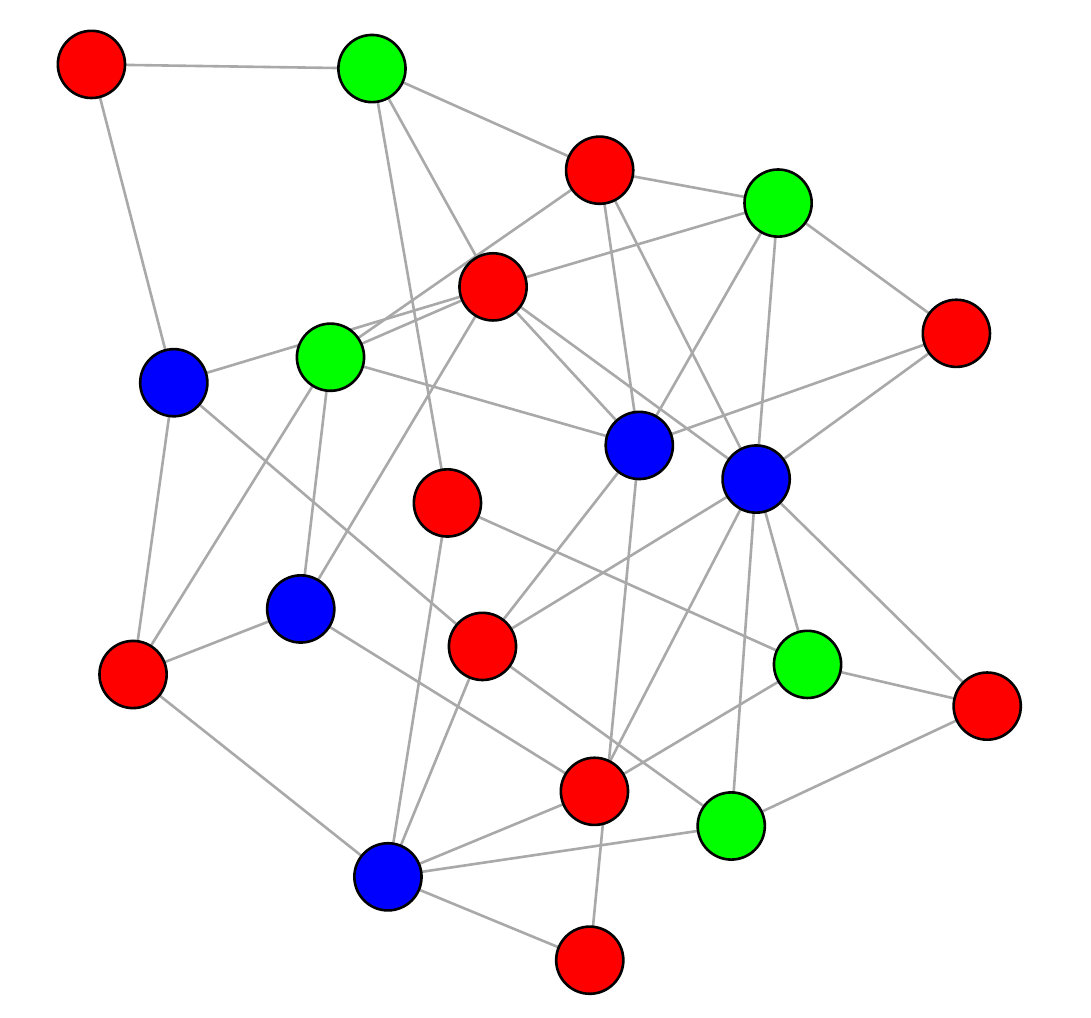}
\end{tabular}
\end{figure*}

\noindent For each edge, once the community co-membership of nodes has been sampled, the sender and recipient can be sampled, incorporating an arbitrary degree distribution into the generative process. For simplicity, the proposed generative model samples senders and receivers uniformly given the set of remaining edges within communities and $\gamma$. The generative process is  shown in Figure \ref{fig:generative process graphical model}. If nodes have a perfect preference for selecting edges within (between) community, then they will only select within (between) community edges until $D = D_M$ at which point the proportion of edges within group will asymptotically approach $D_M$ when $D = 1$. Furthermore, if $\rho = D_M$, the graph will display a constant proportion of within-community edges. Several graphs simulated from the generative process are depicted in Figure \ref{fig:Simulated_Networks}.

%
%

\begin{figure*}[!ht]
\centering
\caption{\label{density_proportion_tradeoff 0 100}
Compartmentalization coefficient $\Upsilon$ values across different $D_M$ -- $D$ combinations. Graphs were simulated from generative process and $\Upsilon$ averaged over 20,000 simulations. The level plots display compartmentalization coefficients recovered from graphs  generated with \textbf{ (a)} : $\rho = 0$,\textbf{ (b)} :  $\rho = 0.5$, \textbf{ (c)} : $\rho = 1$. \bigskip}
\begin{tabular}{ccc}
\textbf{\large(a)} & \textbf{\large(b)} & \textbf{\large(c)}\\
\includegraphics[scale=.37]{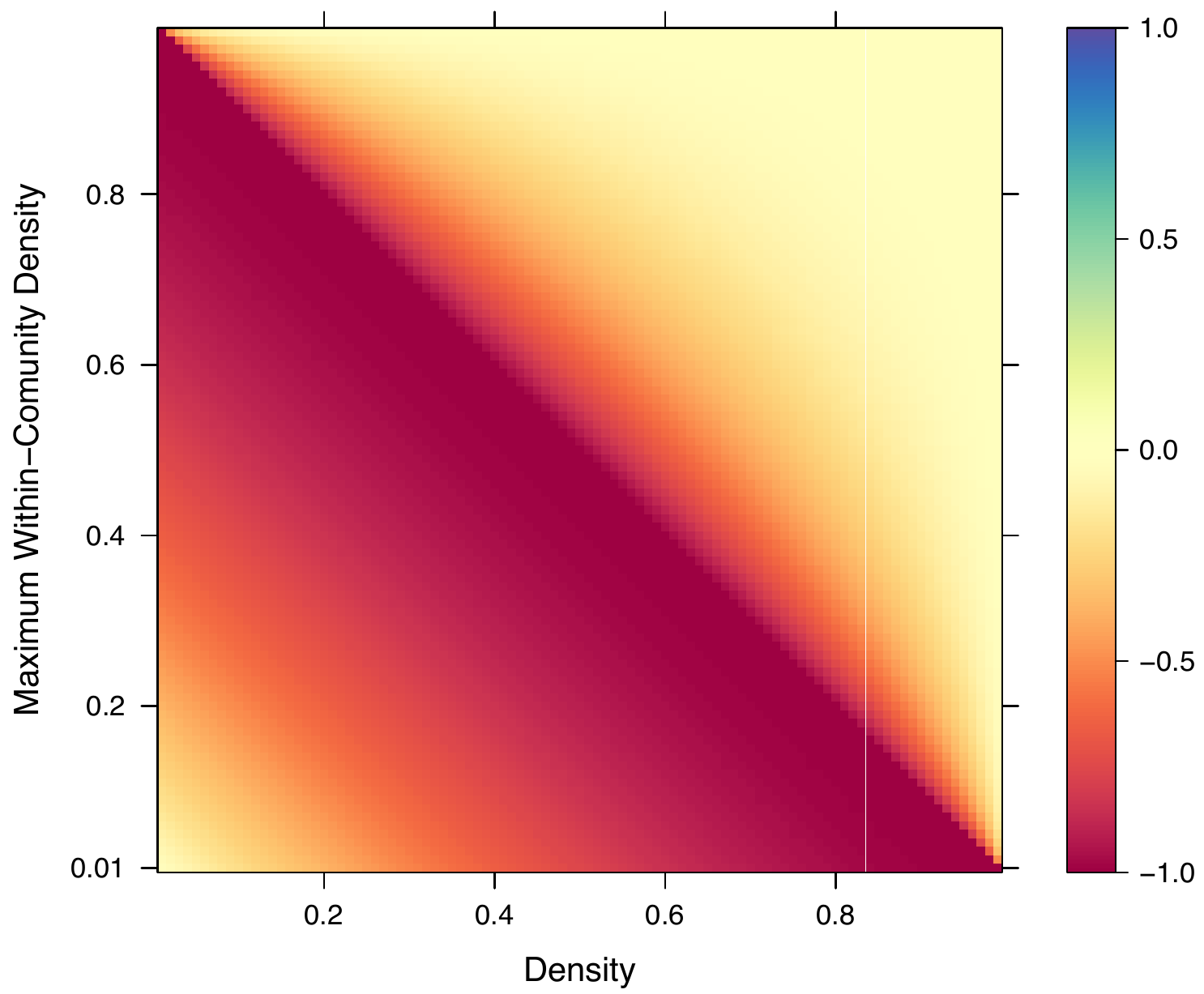}&
\includegraphics[scale=.37]{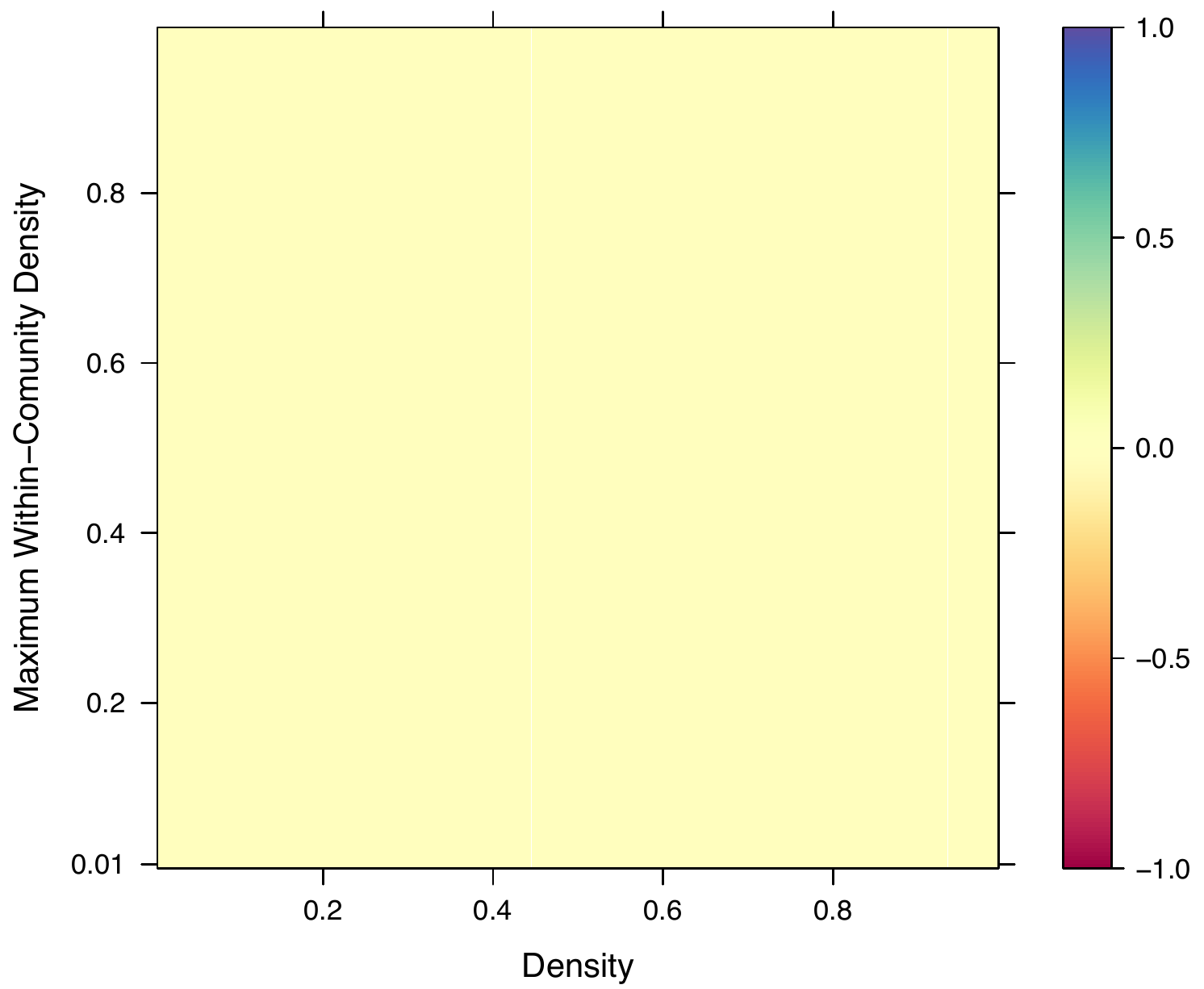}&
\includegraphics[scale=.37]{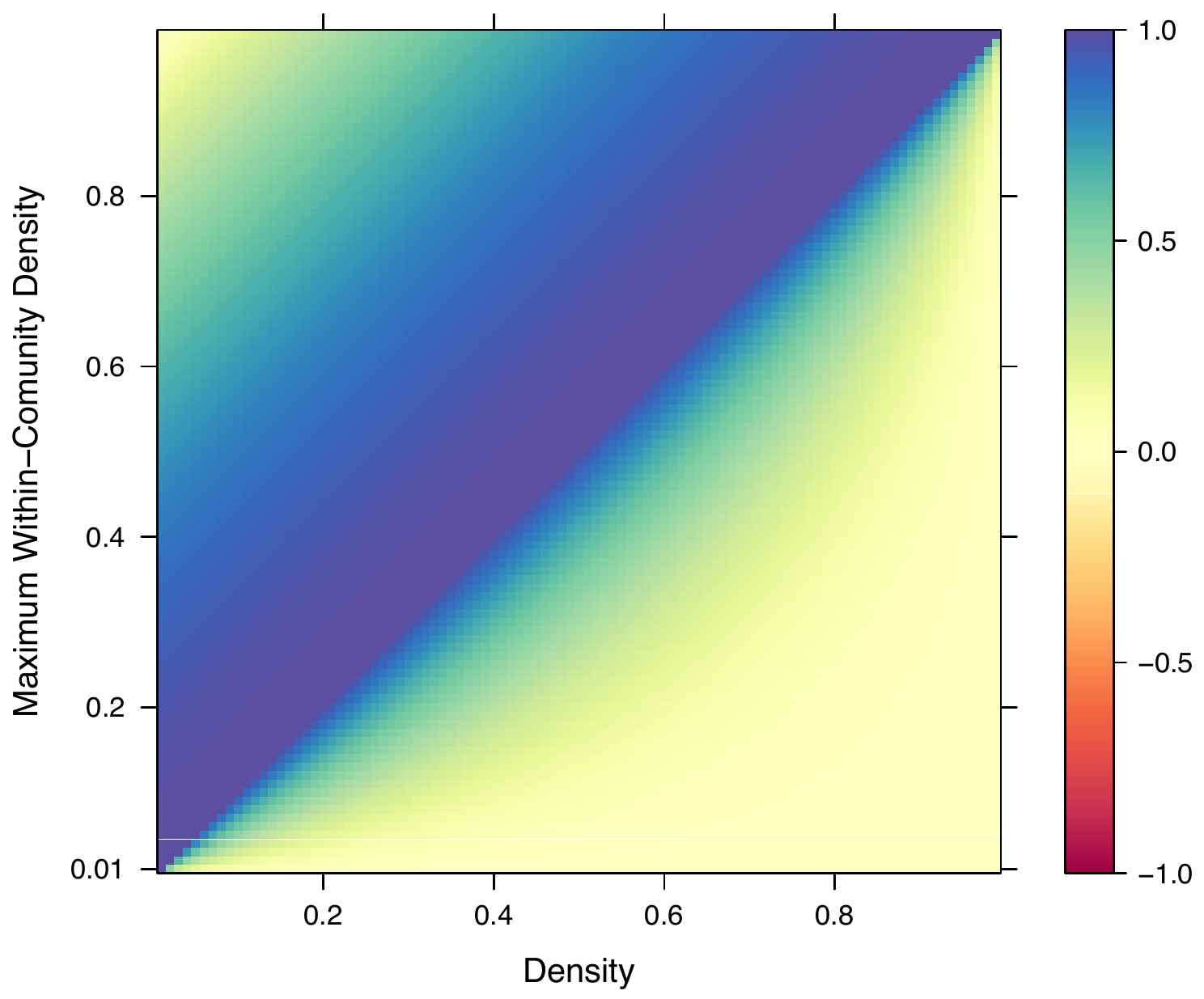}
\end{tabular}
\end{figure*}

To be consistent with the criteria set out for identifying a valid measure of compartmentalization, $\Upsilon$ must be equal to 1 if and only if  $D = D_M$ for a graph generated with $\rho = 1$ (perfect preference for in-community edges so long as they are available). Similarly, this measure must be equal to 0 if and only if  $D = 1 - D_M$ for a graph generated with $\rho = 0$ (perfect preference for out-group edges so long as they are available). As we can see in Figure \ref{density_proportion_tradeoff 0 100} panels \textbf{(a)} and \textbf{(c)}, these criteria are satisfied by $\Upsilon$. 
We can also see from panel \textbf{(b)} in Figure \ref{density_proportion_tradeoff 0 100} that $\Upsilon$ recovers the lack of preference for within or between-community edges when $\rho = 0.5$, essentially serving as a benchmark against which to compare the relative compartmentalization of other graphs.

\subsection{Measure Comparison}

\begin{figure}[!htbp]
\centering
\caption{\label{fig:measure comp}Modularity and Compartmentalization coefficient values for graphs simulated from the generative model with $N = 100$, $D_M = 0.502$ and only one group with more than one member with values averaged over 1000 simulations for $\rho = 0$ (black), \color{red} $\rho = 0.25$ (red), \color{green} $\rho = 0.5$ (green), \color{blue} $\rho = 0.75$ (blue), \color{ProcessBlue} $\rho = 1$ (light blue).}
\includegraphics[scale=.69]{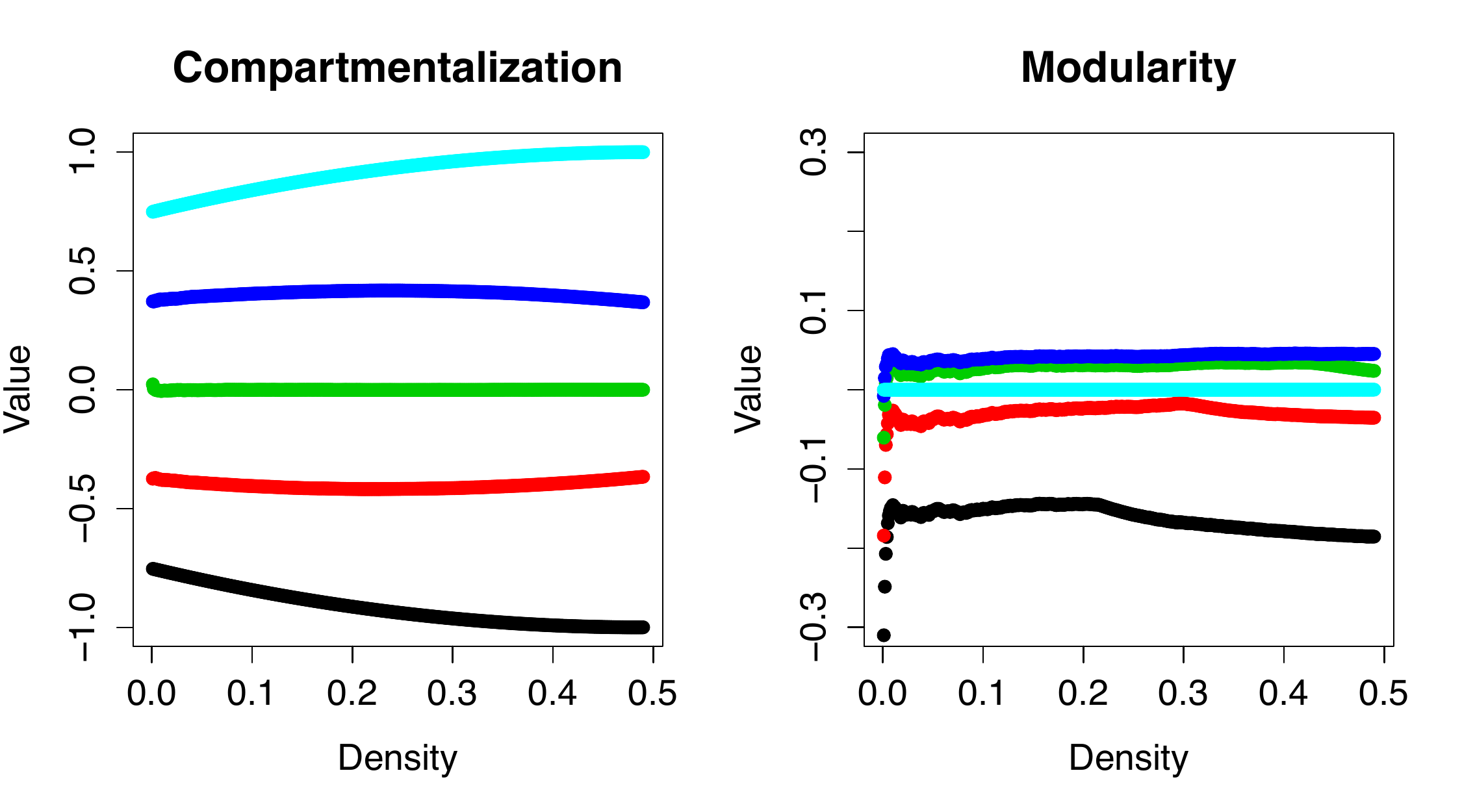}
\end{figure}

One of the most important aspects of the compartmentalization measure introduced in this study is that it is designed to facilitate comparison across graphs.  Figure \ref{fig:measure comp} illustrates the difference between $\Upsilon$ and $Q$ in their validity as a metric of  comparison between graphs with one large group (and all other groups containing only one node) simulated from the generative process described in Figure \ref{fig:generative process graphical model} with varying values of $\rho$. As we can see, the average value of $Q$ across these simulations does not preserve the ordering in compartmentalization implied by the increasing values of $\rho$, while $\Upsilon$ correctly preserves this ordering. 



\section{Application: Political Polarization}
There is a rich literature in political science developing political ideology ideal point estimates for  members of congress based on patterns of roll-call voting on individual bills \citep{poole1997congress, Lewis2004a, Carroll2009}. These ideal point estimates can be leveraged to measure political polarization in congress by tracking the differences in party-mean ideal point estimates over time. While only a small fraction of bills actually make it to a vote, each piece of legislation introduced in congress has a sponsor, who makes an effort to encourage co-sponsorship of the bill by other legislators as a show of support -- with a goal of increasing the likelihood that the bill will advance through the legislative process. A number of recent studies have considered both the act of cosponsorship, and the network of cosponsorship relations between legislators as politically important and providing information beyond roll-call voting patterns \citep{Kessler1996, Fowler2006, Fowler2006a, Harward2010, Kirkland2012}.  Additionally, some authors have sought to advance the measurement and qualification of party polarization in congress using the modularity of co-bill-cosponsorship and co-voting networks \citep{Zhang2008, Waugh2009}. 

\begin{figure}[!htbp]
\centering
\caption{\label{fig:application}Plot of demeaned, standardized, political party modularity and compartmentalization in the Senate co-bill-cosponsorship network and difference in party mean NOMINATE scores from the 96th term of Congress (1979-1980) to the 108th term (2003-2004)}
\includegraphics[scale=.85]{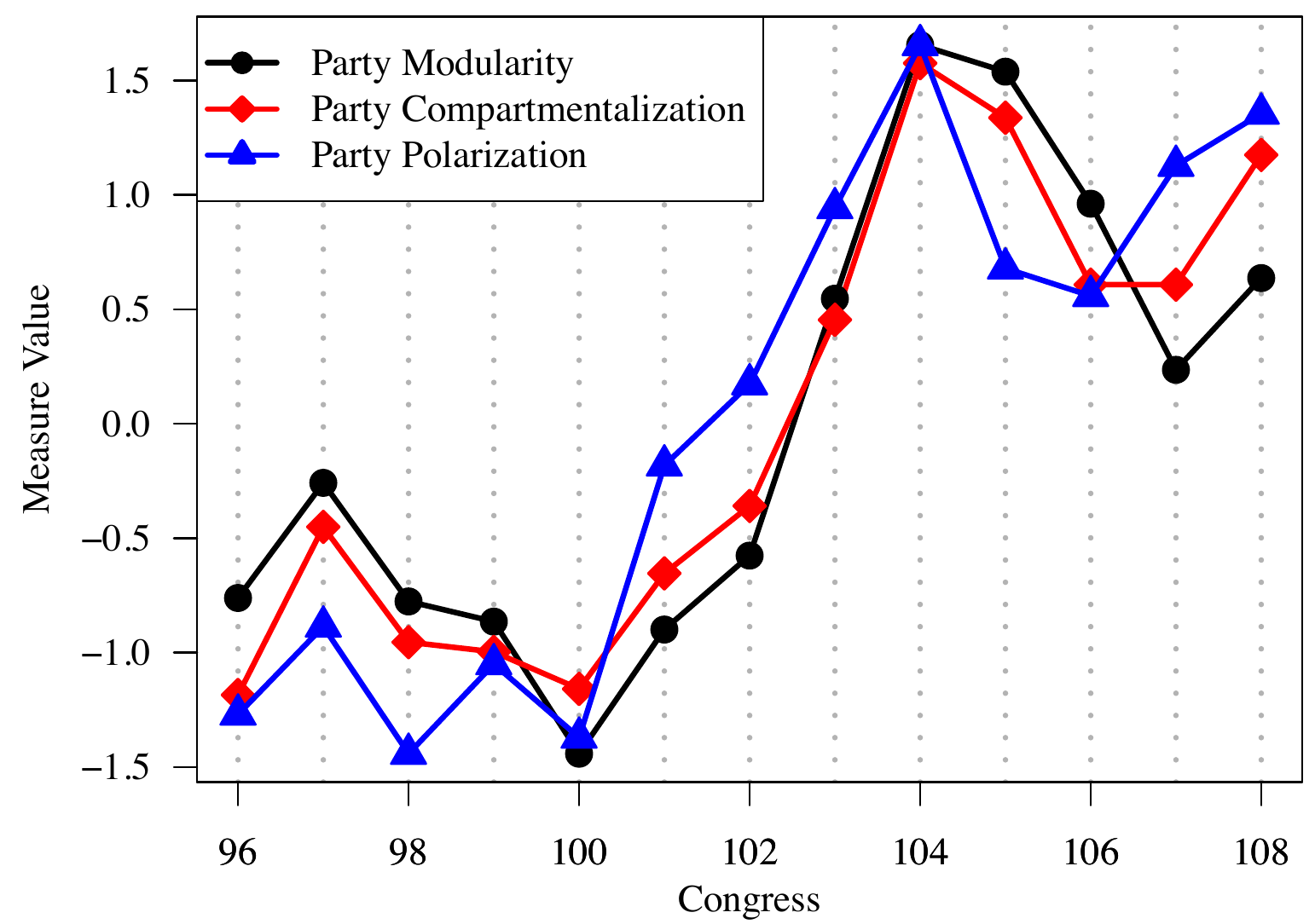}
\end{figure}

This study takes the difference in party-mean ideal point estimates of individual ideology derived from roll call voting as a ground truth measure of polarization and compares modularity and compartmentalization measures against it (Figure \ref{fig:application}). Measures are calculated on the weighted one-mode projection of a cosponsor-bill two mode graph for each session of congress. Weighted graph density is calculated by dividing the sum of weighted ties by the average weighted tie value for present ties times the maximal number of edges possible in the graph. A comparison of correlation coefficients between the two measures and the ground truth measure of ideological polarization using Hotelling's formulation \cite{Hotelling1940} shows a significantly higher correlation between $\Upsilon$ and the ground truth measure than $Q$ and the ground truth measure ($p$ = 0.0345). This application grants further external validity to the new measure of compartmentalization and shows that it can provide improved performance over modularity in measuring polarization on political networks.


%
\section{Conclusions}
The measure of graph compartmentalization proposed in this paper builds on the concept of modularity to allow for principled comparison across graphs of arbitrary structure. The ability to make absolute comparisons about the compartmentalization of graphs has a wide range of applications in social science in the measurement of group separation in observed networks as well as applications in computer science including the comparison of parallel processing problem complexity, for example. 


\bibliographystyle{aipauth4-1} 
\bibliography{library}

\end{document}